\documentclass[conference]{IEEEtran}
\makeatletter
\def\ps@headings{%
\def\@oddhead{\mbox{}\scriptsize\rightmark \hfil \thepage}%
\def\@evenhead{\scriptsize\thepage \hfil \leftmark\mbox{}}%
\def\@oddfoot{}%
\def\@evenfoot{}}
\makeatother
\usepackage{graphicx,epstopdf, amsmath,epsfig,subfigure,endnotes,footnote,multicol,sidecap,url,colortbl,booktabs,algorithm,algorithmic,multirow,enumerate}
\usepackage{amsfonts}
\usepackage [english]{babel}
\usepackage [autostyle, english = american]{csquotes}
\MakeOuterQuote{"}

\usepackage{balance}

\usepackage{amsthm}
\usepackage{mathtools}

\usepackage{environ}
\theoremstyle{plain}
\usepackage{color}
\definecolor{red}{rgb}{1,0,0}

\addto\captionsenglish{\renewcommand{\figurename}{Fig.}}

\NewEnviron{myequation}{%
    \begin{equation}
    \scalebox{0.8}{$\BODY$}
    \end{equation}
    }

\begin{document}
\title{A Review On Game Theory With Smart Grid Security}

\author{
Rahat Masum\\
Department of Computer Science,\\
Tennessee Tech University, Cookeville, TN, USA\\
rahatapu103@gmail.com

}

\maketitle
\begin{abstract}
Smart grid is the modern two way mechanism combining the power grid, control center, smart metering facility, energy routing and customer demand response services. The system being complicated, security vulnerabilities are paramount for the sound operation and process continuation. Since smart grid connects with the end user to the energy providers, these two parties can interact with each other within the whole energy management work flow. In this regard, game theory provides effective insights in the analysis of security measures for smart grid. The mentioned parties will be the players in the game model to provide a solution for the various threats to the grid aspects. In this work, a brief review has presented with the existing approaches to the threat models for divergent sectors of the smart grid. The solution approaches to these threats are based on the game theoretical approaches that connect the attackers and defenders in the scenarios.

\end{abstract}




\section{Introduction}
\label{sec:Intro}
Smart grid revolutionize electricity generation, transmission, and distribution by allowing two-way flows from power generation facilities to end users. A power grid system can, in general, be divided into two main phases: electric power transmission and electric power distribution. Electric power transmission deals with the transmission of the energy generated at the power plants (i.e. the transfer of energy, over transmission lines, to substations that service some geographical areas) whereas power distribution network delivers the electricity to the consumers’ end~\cite{saad2012game}. A smart grid is a large-scale system that extends from a power generation facility to each and every power consuming device such as home appliance, computer, phone, vehicles etc. Thus large-scale nature has increased the possibilities of remote operation of power management and distribution system. With energy being a premium resource, ensuring security against theft, abuse, and malicious activities in a smart grid is of prime concern. For several decades, game theory has been adopted in a wide number of disciplines ranging from economics and politics to psychology~\cite{basar1999dynamic}. Game theory is a mathematical process of modeling the strategic competition between two or more players containing set of specified rules. The final outcome of a game is determined jointly by the strategies chosen by all participants.\\

One of the underlying assumptions in classical game-theoretic designs is that the players are rational. In practical control systems such as the smart grid, as the individual nodes of the system interact and learn their strategies, one or more nodes might deviate from the intended play and make non-rational decisions. In smart grids, it is often possible to enable a limited form of communication between the nodes which paves the way for introducing cooperative game-theoretic approaches with the deployment of advanced networking technologies. In fact, the integration of power, communication, and networking technologies in future grids opens up the door for several prospective applications in which smart grid nodes can cooperate so as to improve the robustness and efficiency of the grid.\\

To provide a comprehensive description of existing game theoretic applications in smart grid networks, in this work, we try to identify key open problems regarding smart grid securities that can be addressed using game theory. The rest of the paper is organized as in Section II, we discuss some aspects of smart grid. Some key game theoretic concepts are overviews in Section III. In Section IV, the survey of game theoretical approaches for different security issues in smart grid is presented.





\section{Smart Grid}
\label{sec:SG}

In order to efficiently deliver power and electricity to their consumers, a composition of intelligent nodes in the power network refers to smart grid. The traditional electrical power grid is unidirectional in nature, where the electricity flows from power generation facilities to end users. However, smart grid can operate, communicate, and interact, autonomously in a two way nature. A smart grid includes diverse and dispersed energy resources and accommodates electric vehicle charging as well. Being a modern electric system, smart grid uses communications, it facilitates connection and integrated operation~\cite{murphy2010enabling}. In the distributed control system, the measurement signals from energy sources (intelligent electric devices, remote terminal units etc.) are communicated to the respective local controller. The local controllers, which communicate among themselves to form a larger intelligent entity, and make decisions on the best possible set of operations to improve the overall performance. Smart grid can be viewed as a combination of Advanced Metering infrastructure (AMI), Supervisory Control and Data Acquisition (SCADA) and Demand Side Management for the energy market.\\

With the combination of sensors, automation and computers this technology improves the flexibility, security, reliability, efficiency, and safety of the electrical system. It also offers consumers increased choice by facilitating opportunities to control their electricity use and respond to electricity price changes by adjusting their consumption. It is ensured that implementing two-way communications in the smart grid will not only allow dynamic monitoring of the use of electricity but also open up possibilities of automated scheduling of electricity use~\cite{doe2010communications}.


\section{Game Theory}
\label{sec:GT}
Game theory is a mathematical framework that can be divided into two main branches: noncooperative and cooperative game theory. Noncooperative game theory can be used to analyze the strategic decision making processes of a number of independent entities. The entities can be known as players, and have partially or totally conflicting interests over the outcome of a decision process. The process is affected by their actions. However, the term non-cooperative does not always imply that the players do not cooperate as in person, but it means that, any cooperation that exists must be self-induced with no communication or overlapping of strategic choices among the players. Noncooperative games can be seen as capturing a distributed decision making process that allows the players to optimize, without any coordination or communication, objective functions coupled in the actions of the involved players assuming players are unable to coordinate or communicate with one another directly. However, for games in which the players are allowed to communicate and to receive side payments, it may be of interest to adopt fully cooperative manners. Cooperative games allow to investigate how one can provide an incentive for independent decision makers to act together as one entity. The approach is to improve their position in the game. Players can cooperate with each other but each want to maximize their benefit, thus, conform with some notion of equilibrium. The strategies can be:

\begin{itemize}
\item \textbf{Simultaneous}: making moves without knowing other’s decision.
\item \textbf{Sequential}: having knowledge of the moves by before acting.
\item \textbf{Zero sum}: gain of one is loss of another player.
\item \textbf{Nash Equilibrium}: one can not go further with own interest if other does not change their position.
\end{itemize}

Noncooperative games can be grouped into two categories: static games and dynamic games whereas cooperative game theory encompasses two parts: Nash bargaining and coalitional game. In static games, the notions of time or information do not affect the action choices of the players. Thus, in a static setting, a noncooperative game can be seen as a one-shot process. In the process the players take their actions only once. This may be simultaneously or at different points in time. In contrast, the players have some information about each others’ choices in dynamic games. Players can act more than once, and time has a central role in the decision making. One of the most important solution concepts for game theory is that of a Nash equilibrium. The Nash equilibrium characterizes a state in which no player can improve its utility by changing unilaterally its strategy, given that the strategies of the other players are fixed. These exists another theoretical concept of auction theory. It is used to study the interactions between a number of sellers, each of which has some commodity or good to sell, and a number of buyers interested in obtaining the good. The outcome of the auction is the price at which the trade takes place as well as the amount of good sold to each buyer. In the smart grid model, a micro-grid acts as a seller or a buyer is dependent on its current generation and demand state. The strategies of every member in S correspond to the price at which it is willing to buy/sell energy and the quantity that it wishes to sell/buy. The objective of each player is to determine the optimal quantity and price at which it wants to trade. Game theory is treated as a key analytical tool in the design of the security problems in the smart grid.\\

\section{Smart Grid Security}
\label{sec:SGS}
A smart grid comprises four major facilities: generation, transmission, distribution, and end user consumption. The ability to monitor and influence each user’s usage in real time can enable distribution operators or utilities match supply with demand effectively and realize the potential of digital power~\cite{lim2016multi}. The challenges of ensuring security in a smart grid are vast and widely distributed. This is due to the nature of diversity of the components and the contexts where smart grids are deployed. The cybersecurity objectives can be classified into the following three categories known as CIA:
\begin{itemize}
\item \textbf{Confidentiality}: Protecting privacy and proprietary information by authorized restrictions on information access and disclosure about any consumer's energy consumption pattern, usage data, billing etc.
\item \textbf{Integrity}: Protecting against the unauthorized modification or destruction of information Unauthorized information access opens the door for mishandling of information, leading to mismanagement or misuse of power causing alteration in the system model and procedures.
\item \textbf{Availability}: Ensuring timely and reliable access to information and services. It can be compromised by disruption of access to information which undermines the power delivery.
\end{itemize}
Smart grid security mechanism should be enforced at several layers including physical and logical layers. Privacy preserving Smart Metering through the information network in a smart grid frequently transports confidential information relating to customers, for example, identity, location, possession of electronic appliances and devices, and power usage profile~\cite{yang2016false}. In this regard, data encryption in smart grid, from meter to utility center, is a useful tool to prevent threats in the system. It will help to preserve the confidentiality of data. All smart grid devices, for example, meters, collectors, processors, and routers, must be enabled with encryption processing capabilities~\cite{hayden2010there}. Another common concept of security mechanism is authentication. It is the process of determining that a user or entity is ensured as the same been claimed. Smart grid applications must have strong authentication capabilities, to detect and reject unauthorized connections between its components, for example, meter and the utility interfaces~\cite{li2010secure}.\\

In this work we are providing a brief discussion of security issues in smart grid using the numerous game theoretical approaches presented by different analyzers. For the convenience of discussion, existing works have grouped into three subsections. Each section contains the problem domain and solution approach by game theory in the specified direction.\\

\begin{table}[h]
\begin{tabular}{p{0.2\linewidth}p{0.25\linewidth}p{0.28\linewidth}}
\hline
\textbf{Author} & \textbf{Game theory approach} & \textbf{Smart Grid Area}\\ \hline
Wang \textit{et al.} & Bayesian honeypot game
 & Advanced Metering Infrastructure \\ \hline
Abercrombie \textit{et al.} & Agent Based Game Theory & Advanced Metering Infrastructure \\ \hline
Ismail \textit{et al.} & Stackleberg game & Advanced Metering Infrastructure \\ \hline
Cardenas \textit{et al.} & Noncooprative game & Advanced Metering Infrastructure \\ \hline
Ma \textit{et al.} & Markov game & Advanced Metering Infrastructure  \\ \hline
Hewett \textit{et al.} & Noncooprative game & Supervisory Control and Data Acquisition \\ \hline
Hong \textit{et al.} & Noncooprative game & Supervisory Control and Data Acquisition \\ \hline
Law \textit{et al.} & Stochastic security game & Supervisory Control and Data Acquisition \\ \hline
Nguyen \textit{et al.} & Coalitional game & Supervisory Control and Data Acquisition \\ \hline
Soliman \textit{et al.} & Cooperative game & Demand Side Management \\ \hline
Belhaiza \textit{et al.} & Noncooperative & Demand Side Management \\ \hline

\end{tabular}
\end{table}

\subsection{Advanced Metering Infrastructure (AMI)}
AMI is the deployment of interconnected smart meters that enable two-way communication in order to have continuous and timely monitoring of meter data, outage reporting, and service connect/disconnect. AMI consists of smart meters, data aggregators, central system (AMI head end), meter data management system (MDMS), and the communication networks and enabling communication technologies. Apart from measuring power consumption, smart meters can also monitor statistical data, and report to the consumers.\\

Among various types of threats in the smart grid, distributed denial-of-service (DDoS) is a typical attack that severely threatens availability of the communication network resources. DDoS attack refers to any event that can reduce or eliminate the proper execution of the network. Wang \textit{et al.} focus on distributed denial of service (DDoS) attacks in the AMI network. They introduce honeypots into the AMI network as a decoy system to detect and gather attack information. Honeypots are security resources that help attract, detect, and gather attack information. By pretending to be normal servers to attract the attackers, honeypots can consume attackers’ resources and time. They can also influence and interfere with the choice of intruders, and further detect the intruders’ attack intention. Other than production systems, the main system can monitor any suspicious intrusion to honeypots. The authors analysis over the interactions which happens between the attackers and the defenders help them derive an optimal strategies for both players~\cite{wang2017strategic}. In the work, dynamic attack is introduced to push invalid data into a system, and/or to damage or destroy data already stored in it. The attacker can first utilize an anti-honeypot to detect the honeypot proxy server in the target network by transmitting initiative packets. Once the honeypot server discovered, the attacker can bypass the honeypot, and access to the target network through other channels. The work presented introduce a Bayesian honeypot game model to derive and prove that the equilibrium conditions can be achieved between legitimate users and attackers, for the strategies of honeypots and anti-honeypots. One example scenario considered targeting a critical server (e.g., a FTP server or a Web server) in an AMI network which may lead to network paralysis, power shortages, and power overload in the smart grid~\cite{wang2017strategic}. Honeypots can be set at the other side of the firewall away from the energy provider. They propose to analyze the interactions between attackers and defenders to derive optimal strategies for honeypots deployment in the AMI networks. By using \textit{Bayesian Nash Equilibrium}, the work provides effectiveness of defense in the game model’s detection rate between 70\% and 85\% to  deploy honeypots in AMI network, and add the decoy performance. These results indicate that increasing the number of honeypots in the AMI network can significantly reduce the anti-honeypot accounts for the proportion of the total number of servers.

About the CIA issues in AMI one Dynamic Agent Based Game Theory (ABGT) has been presented by Abercrombie \textit{et al.}. The work select some failure scenarios from the set of cyber security and impact analyses for the electric sector developed by The National Electric Sector Cybersecurity Organization Resource (NESCOR) Technical Working Group 1 (TWG1). One demonstration about how to model the AMI functional domain using a set of rationalized game theoretic rules has decomposed from the failure scenarios. In terms of those scenarios simulation approach has been used~\cite{abercrombie2014security} over the AMI network with respect to CIA. Dynamic Agent Based Game Theoretic (ABGT). The authors develop each ABGT simulation taking as input a model of a specific failure scenario of interest. Then the interactions between an attacker and the administrator (defender) were modeled as a two player stochastic game for which best-response strategies (Nash Equilibriums) were computed. Depending on the scenario, the attacker executes one of many actions with	probability of deciding to do the action and, probability that the action will be successful. As well, the defender performs actions, by the probability of detecting that something is wrong or inconsistent with the normal state of operation. Providing the normal AMI states are known, the simulation proposed by the authors, will try to limit the defender’s actions, which is a counter action to the most current action performed by the attacker. The game is formulated in a way that before the defender performs any counter action, a detection action is required to confirm the type of attack. The test case simulation represents time unit as one minute and within each time unit, the simulator thread visits each agent giving them the opportunity to perform an action or not. The result discussed in the paper has the confidentiality variation for the attacker/defender interplay over time (up to 190 minutes) and the attacker with the highest arrival rate produces the greatest gains~\cite{abercrombie2014security}. On the other side of the game, an administrator can suffer damages that result in system downtime or theft of customer data. The future work is directed to handle situation with more than one attacker per network and more than one administrator managing the network at the same time. The contribution can be identified as the defender being able to more realistically determine how much security is needed in the AMI from both the utility’s and customer’s perspective.\\

A good approach to the confidentiality of information in the AMI consisting of nodes with interdependent correlated security assets are discussed in the work of Ismail \textit{et al.} ~\cite{ismail2014game}. Their main concern is about providing a game model regarding-
\begin{itemize}
\item How attacker chooses their targets to collect the maximum amount of data on consumers.
\item How defender chooses the encryption level of outbound data on each device in the AMI.
\end{itemize}
The authors formulates the problem as a noncooperative game and analyze the behavior of the attacker and the defender at the Nash equilibrium. Their approach stands over the concept of on each node, the defender an encryption level of outbound data on each node and data on each communication link is encrypted with different encryption keys or using different encryption algorithms. Since attacker tries to intercept data by attacking the nodes without being detected, if the attacker wants to intercept data sent by node \textit{i}, he can either attack node \textit{i} or attack the parent node of \textit{i}. For the defender, on each node, an Intrusion Detection System (IDS) is installed with a detection rate. The defense mechanisms deployed to protect a device depend on the value of data generated, stored, or processed by that device. Utility functions are defined as $U_A$ for attacker to max p(s) =$U_A$(p,s) and $U_D$ for defender to max s(p) = $U_D$(p(s),s). Using Stackleberg game approach this work solves the issue with a leader who chooses his strategy first. The follower, notified by the leader’s choice, chooses his strategy. The leader tries to anticipate the follower’s response and chooses the strategy that yields the maximum payoff knowing what will be the reaction of the follower.\\

A very common problem in the smart grid is electricity theft. With the development of advanced metering infrastructure in smart grid, more complicated situation in energy theft has emerged and many new technologies are adopted to try to solve this problem. About the energy theft Jiang \textit{et al.} have investigated the system model and security requirements of AMI in smart grid presenting an attack tree based threat model for AMI. We further categorize the energy-theft detection schemes in AMI. The authors also discussed we discuss the challenging issues in energy theft detection and provide some research directions~\cite{jiang2014energy}.Attackers in AMI can be classified as : curious eavesdroppers who are only interested in the activities of their neighbors, greedy customers who want to crack the AMI in order to steal electricity, malicious eavesdroppers collecting metering data for some vicious purposes such as house breaking or intrusive data management agencies who want to collect customers’ private information for marketing or economic purposes.\\

Cardenas \textit{et al.} shows one anomaly detection as a game between the electric utility and the electricity thief~\cite{cardenas2012game}. The game formulated as a thief (fraudulent consumer) will steal a predefined amount of electricity while minimizing the likelihood of being detected and the electric utility wants to maximize the probability of detection and minimize the operational cost for managing this anomaly detection mechanism. Their solution talks about Thief Consumer stealing electricity is a subset from the total set of customers . To find the Nash equilibrium of game consumers are assumed independent. An electric distribution utility has an AMI deployment collecting a time-series of electric power consumption $y_{ik}$. For every time step \textit{k} and every customer $q_i$ being the expected total consumption of user and $q_{iU}$ being the expected unbilled part of the consumption of user $q_{iU}=0$ for honest users. The revenue of a distribution utility is the sum of the tariffs from all customers adding  the recovered fines from the detected electricity theft subtracting the investment in protecting infrastructure against electricity theft.

A centralized meter data management (MDM) solution  performs analysis of the time series received by consumers comparing them to historical trends  and correlates them to other customers in similar residences or businesses. The operational cost of managing the anomaly detector is quantified by the resources/efforts the distribution utility assigns for dealing with false alarms by the proposed solution~\cite{cardenas2012game}.\\

Simultaneous attack on smart meters to destabilize the grid could be done through cyber means. These attacks cannot be determined by physical attacks alone. The interactions between the providers and attackers of the smart grid and their optimal strategies can be modeled as a Markov game ~\cite{ma2013scalable}. In their work authors define the problem of smart grid protection against cyber-space attacks which can be on the communication network. Their consideration is for a decentralized real-time energy market enabled by a smart meter infrastructure (AMI). This spans distributed consumer sites and support real-time Demand Response (DR). Usually, the smart meters report usage information periodically allowing their owners to bid for different levels of power supply in real time from distributed generation. Based on the bidding results, the DR functions in the smart meters
control the operations of their attached loads. These loads can be modeled as smart appliances for houses or smart vehicles. A game model describes the defense activation status of links which relay bids between the smart meters and the bidding market. In each time step, the players choose a pair of actions which, together with some underlying probabilistic physical events, may cause state transitions in a Markov manner. The work was mainly focused onthe computation of equilibrium best policies for both players~\cite{ma2013scalable}. The policies estimate optimal mixed strategies of the player concerned, and an optimal strategy is one that maximizes the minimum reward which is under the best strategy of the opponent in the lone term manner.\\

\subsection{Supervisory Control and Data Acquisition (Scada)}
The cyber and communication components of electrical grids are formed by the Supervisory Control and Data Acquisition (SCADA) systems. Operators use SCADA systems to receive data from and send control signals to grid devices. This distributed system helps to make physical changes that benefit grid security and operation
~\cite{berthier2012state}. If a SCADA system is compromised by a cyber attack, the attacker may alter these control signals with the intention of degrading operations or causing widespread damage to the physical infrastructure~\cite{backhaus2013cyber}. In general, the attacker has advantages over the defender in that his attack action takes relatively little effort and though results in a small direct payoff to him, it results in a huge loss to the defender.\\

In the context of cyber-security of smart grid SCADA systems, the game theory can work same as before, played between an attacker and a defender. Hewett \textit{et al.} defined a two-player game on the smart grid SCADA system representing an attacker and a defender (or SCADA security administrator). Each player has a finite set of actions. Certain cyber security attacks require more than one action step and as the attack is advanced, the payoffs change. The authors defines the payoffs to the attacker should increase while payoffs to the defender should escalate into a huge loss. To model such behavior, the player's payoff in a current state or decision node in the game tree must depend on his payoff in a previous state parent node of the decision node as mentioned~\cite{hewett2014cyber}. To predict how system design choices affect the outcome of attacker-defender interactions, a description of when player decisions are made and how these decisions affect the system state, i.e. a “game” definition must be analyzed. Sophisticated attacker strategies may be carried out over many time steps.\\

In their work Hong \textit{et al.} finds the optimal energy route to meet the electricity demand for all nodes in an smart grid network. They divide it into subproblems as:
\begin{itemize}
\item For each selling/buying node to find the proper price to maximize its own profit.
\item For each node to determine the optimal power transmission path according to a potential counter party
\end{itemize}
After solving the fist subproblem, the transaction price at time period is obtained. The amount of electricity sold by each supply node, and the amount of electricity purchased by each demand node utilizes the results of the fist subproblem~\cite{hong2016game}. This work proposed three strategies for determining the desired prices of the supply and demand nodes. Firstly, as in optimistic strategy for the desired trading price, to maximize the profit, rather than the possibility of the sale/purchase. Secondly, in the medium strategy player pursue the maximization of both the possibility of a sale or purchase and profits from trading. Finally, pessimistic strategy maximize the possibility of a sale or purchase, rather than a profit, when predicting a quite different determination of price.\\

Data injection attack consequences are quantified using a risk assessment process where Law \textit{et al.} provides a well-known conditional value at-risk (CVaR) measure to estimate the defender’s loss due to load shed in simulated scenarios. The work calculates the risks to incorporate into a stochastic security game model as input parameters~\cite{law2015security}. The final decisions on defensive measures are obtained by solving the game by the proposed dynamic programming techniques which take into account resource constraints. Hence, formulated security game provides an analytical framework for choosing the best response strategies against attackers and minimizing potential risks presented in the paper. According to the solution,security games provide an analytical framework for modeling the interaction between malicious attackers who aim to compromise a smart grid, and operators defending it. A numerical study illustrating the framework has been developed, and it demonstrate the dependence of the optimal strategy on the employed risk measure.\\

\subsection{Demand Side Management}

Demand management can be treated as a form of indirect generation. Smart metering, with two-way communications capability allow utilities to collect and analyze usage information at narrow interval rather than relying on manual monthly reading. Classical demand-side management schemes such as direct load control and smart pricing are focused on the interactions between a utility company and each individual end-user. The utility companies have to consider if they have still a security program to upgrade after a gap analysis or have no security program yet so an exhaustive adaptation of the frameworks is needed~\cite{pallotti2011smart}.\\

Demand-side management (DSM) can be classified into: reduce consumption, and shift consumption. Methodologies for deriving reliability performance of wireless communication networks support demand response control. The use of cooperative games for enabling a coordinated demand side management among the users. It can, subsequently, lead to a more efficient load distribution and less costs on the utility operator. A simple class of congestion games, as a means for performing dynamic pricing so as to control the power demand in an effort to achieve, not only provide net energy savings, but also an efficient utilization of the energy. Instead of focusing only on the individual user consumption such as in classical schemes, it is better to develop a demand-side management approach that optimizes the properties of the aggregate load of the users. This is enabled by the deployment of communication technologies that allow the users to coordinate their energy usage, when this is beneficial. As a noncooperative game for scheduling appliances a demand-side management scheme enables to schedule the shiftable appliances while minimizing the overall energy consumption and, thus, the charges on the consumers.\\

An approach based on a coalitional game is proposed to integrate and resolve the conflicting interests by Nguyen \textit{et al.}~\cite{nguyen2013game}. They uses microgrid with three players to explain problem by proposing an optimal strategy for integration and arbitration of conflicting interests. This can be between producer or consumer agents and network agents. In the work, cooperative game theory is used together with agent-based techniques to optimally allocate resources among involved actors. These three techniques combines to enable an advanced control layer at distribution level, integrate decentralized network functions and real-time market service to empower local actors with appropriate incentives, and solve the cooperation of functions by multi-objective optimization. Th authors introduce agent-based functions of power routing and power matching With direct control actions, the function of power routing can respond to any significant change in the power grid within a very short time after the action of protection equipment. For each network agent representing a local area network (LAN), three functional operation is handled which enables management, coordination, and execution of actions of the active parties within their areas. The proposed work address issues mentioned as: Conflicting Interests, Provision of Capacity for Ancillary Services, and  Momentary Marginal Cost for Power Routing. The game is created with a pair when is the set of players (agents), and is a function that assigns for every coalition a real number representing the total benefit achieved by. The coalitional game starts from the results of the matching progress and requests of ancillary services from network operator.\\

In the Demand Side Management Soliman \textit{et al.} discussed the problem with customers equipped with energy storage devices. They talks about two game approaches as cooperative and Stackelberg game model. Cooperative one is played between the residential energy consumers, while the second one is played between the utility provider and the energy consumers. In the Stackelberg game, the utility provider sets the prices to maximize its profit knowing that users will respond by minimizing their cost. The authors present a new cost function applicable to the case of users selling back stored energy. Stackelberg game is shown to be the general case of the minimum Peak-to-Average power ratio (PAR) problem. Centralized and distributed algorithms are presented to solve the game. The distributed algorithms has the proof of heir convergence within both player interaction. The solution work  presents an extensive set of simulation results covering a range of possible scenarios. The contribution can be mentioned as finding the improved collaboration between the utility and users results in lower user cost and lower system PAR~\cite{soliman2014game}.\\ 

Nguyen \textit{et al.} talks about the noncooperative game theoretical frmework to model DSM problem with energy storage devices. The proposed algorithm discusses to reduce the total energy cost of the system compared with the system without integrating energy storage devices for users~\cite{nguyen2012demand}. This algorithm requires each user communicates with the energy service provider. However, one limitation about the algorithm mentioned in the work is not to converge to the optimal solution in the centralized design. Based on the definition of the payoff and strategies in the game, the users try to select their energy consumption and battery charging or discharging schedule to minimize their energy cost and a user's choice to maximize its own payoff function assumes all other strategies fixed.\\

Reliability a sustainability of the electricity grid can also be improved through customer’s involvement in decisions about his energy consumption, adjusting both the timing and quantity of his electricity use. Shifting customer load during periods of high demand to offpeak periods flattens the load curve, saving the building of additional generation capacity. This data can be used by utilities to verify impact of new planning strategies or build targeted programs to appeal to specific segments of customers. Regarding this, Ibars \textit{et al.} propose distributed load management in smart grid infrastructures to control the power demand at peak hours, by means of dynamic pricing strategies~\cite{ibars2010distributed}. Proposed congestion game model was simulated and shown to reach an equilibrium solution. The solution determines both the user demand vectors and prices paid. This equilibrium point being a local optimum for the selfishly acting user, is also a global solution to the network problem was concluded by the authors. The work presents the game model where the players are the end customers. The active set of strategies is the distribution of the demand across the day, and the cost function is the players’ aim at minimizing is the price they will be charged by the provider. In this situation players should take into account their own preferences. A congestion game strategy to address the problem of load control by implementing a distributed end-user energy consumption scheduling provides network competition. The resulting cost is a function of the level of congestion.\\

Belhaiza \textit{et al.} propose a noncooperative game theoretic model for the management of smart grid’s demand considering the packet error rate. In this paper, a queuing model is used which tries to quantify the packet loss resulting from congestion at data aggregation unit (DAU)~\cite{belhaiza2015game}. To design a control strategy for the demand response from energy retailers for neighborhood-area-network (NAN) customers, a 0-1 mixed linear programming approach to compute nondominated extreme Nash equilibrium using noncooperative multiagent game theoretic model is proposed. The work introduces a \\textit{($n + p$)}-person noncooperative game where each agent aims at maximizing his own utility depending on a number of conditions. The conditions for user and providers are mentioned. For user, total amount of energy requested from all providers cannot exceed the demand and also the average amount of energy requested from each provider cannot exceed the amount of energy supplied. For provider, condition holds as total amount of energy supplied to all users cannot exceed their own (production and delivery) capacity and the amount of energy supplied by provider to each user cannot exceed the average amount of energy requested.\\

The theoretical amount represents the amount of the requests sent to the provider, while the real amount represents the amount of the requests received by the provider. Due to the average packet error rate $p_{ij}$ measured experimentally during the transmission of the information from user to provider over the wireless network deployed in the NAN, the authors define a Nash equilibrium as a situation where users and providers maximize simultaneously their individual utility functions. In the equilibrium, each user node may represent a group of homes connected via one controller and providers can possibly exchange information about their loads. The payoffs will be set to negative for any user would prefer delaying its demand. Also, any provider would prefer not responding to any request which generates a negative payoff in the game. The authors optimize the transmission rate from the DAU to minimize the impact of packet loss.\\

\section{Conclusion}
\label{sec:conclusion}
Smart grid being a widely distributed engineering system, falls victim to numerous security issues both from physical and cyber means. Different components in the smart grid architecture introduce different sorts of attack scenarios and threats. Game theory being a mathematical model of analysis works as an analysis approach to many of these security vulnerabilities. In this literature survey, we distributed the solution approaches to specific areas of smart grid using the existing game theoretical methodologies. We also discussed the contributions and limitations of the approaches to some extent. Out future work will be based on the limitations on the present situations for the security measures. Game theory will be one of the solution methods as per our findings throughout this review. With the proper identification of problem domain and acting user, threat model analysis will be prolific to build a stable and secure environment for smart grid.


\balance
\bibliographystyle{unsrt}
\bibliography{References}

\end{document}